\documentclass[12pt]{article}
\def\int {\intop \limits}

\def\fnote#1{\footnote}
\begin{document}

\newcommand{\dst}[1]{\displaystyle{#1}}
\newcommand{\barl}{\begin{array}{rl}}
\newcommand{\ball}{\begin{array}{ll}}
\newcommand{\ear}{\end{array}}
\newcommand{\barc}{\begin{array}{c}}
\newcommand{\e}{\mbox{${\bf e}$}}
\newcommand{\J}{\mbox{${\bf J}$}}
\newcommand{\be}{\begin{equation}}
\newcommand{\ee}{\end{equation}}
\newcommand{\aq}[1]{\label{#1}}
\renewcommand \theequation{\thesection.\arabic{equation}}

\title{Anomalous magnetic moment of the electron in a medium}
\author{V. N. Baier
and V. M. Katkov\\
Budker Institute of Nuclear Physics,\\ Novosibirsk, 630090,
Russia}

\maketitle

\begin{abstract}
Radiative effects are considered for an electron moving in a
medium in the presence of an external electromagnetic field.
Anomalous magnetic moment (AMM) of the electron in $\alpha$-order
is calculated under these conditions in the form of
two-dimensional integral. Behavior of AMM of high energy electron
under influence of multiple scattering in a medium is analyzed.
Both mentioned effects lead to reduction of AMM.
\end{abstract}

\newpage
\section{Introduction}

The contributions of higher orders of the perturbation theory
over the interaction with an electromagnetic field give the
electromagnetic radiative corrections to the electron mass and
lead to appearance of the anomalous magnetic moment (AMM) of the
electron \cite{S}. It is known that under influence of an external
electromagnetic field these effects in particular the AMM of
electron are changed essentially \cite{BKS1}, \cite{BKF}. Here we
consider how the mentioned effects modify in a medium.

The correction to the electron mass is described by the amplitude
defined by the diagram of the electron self-energy. We use the
operator quasiclassical method (see \cite{BKF}, \cite{BLP},
\cite{BKS}). In this method the mentioned amplitude is described
by the diagram where the electron first radiates a photon passing
to a virtual state and then absorbs it. This corresponds to use
of the non-covariant perturbation theory where in the high energy
region only the contribution of this diagram survives. For the
electron with energy $\varepsilon \gg m~(m$ is the electron mass)
this process occurs in a rather long time (or at a rather long
distance) known as the lifetime of the virtual state
\begin{equation}
l_f=\frac{2\varepsilon}{q_c^2},
\label{1.1}
\end{equation}
where $q_c \geq m$ is the characteristic transverse momentum of
the process, the system $\hbar=c=1$ is used. When the virtual
electron is moving in a medium it scatters on atoms. The mean
square of momentum transfer to the electron from a medium on the
distance $l_f$ is
\begin{equation}
q_s^2=4\pi Z^2\alpha^2n_a\ln(q_c^2 a^2)l_f,
\label{1.2}
\end{equation}
where $\alpha=e^2=1/137$, $Z$ is the charge of nucleus, $n_a$ is
the number density of atoms in the medium,  $a$ is the screening
radius of atom.

In the case of weak scattering $q_s \equiv \sqrt{q_s^2} \ll m$
the influence of a medium is weak, in this case $q_c=m$. At high
energy it is possible that $q_s \geq m$.In this case the
characteristic value of the momentum transfer (giving the main
contribution into the spectral probability) is defined by the
value of $q_c$. The self-consistency condition is
\begin{equation}
q_c^2=q_s^2=\frac{8\pi \varepsilon Z^2\alpha^2n_a\ln(q_c^2
a^2)}{q_c^2} \geq m.
\label{1.3}
\end{equation}
With $q_c$ increase the lifetime of the virtual state \ref{1.1}
decreases and correspondingly the corrections to the electron
mass and AMM of the electron are suppressed.

In the present paper we consider a motion of the charged particle
in a medium in presence of an external electromagnetic field. In
Sec.2 the general expressions for the correction to the electron
mass and for AMM of the electron are derived to order $\alpha$ of
the perturbation theory. In Sec.3 we analyze the influence of
medium on AMM of the electron in detail.

\section{Radiative corrections to the electron mass}
\setcounter{equation}{0}

In the frame of the operator quasiclassical method the probability of
photon emission which will be used below is
\begin{equation}
dw=\frac{\alpha}{(2\pi)^2} \frac{d^3k}{\omega}
\int_{}^{}dt_1dt_2 R^{\ast}(t_2)R(t_1)
\exp\left\{-i\frac{\varepsilon}{\varepsilon-\omega}[kx(t_2)-
kx(t_1)] \right\},
\label{2.0}
\end{equation}
where $k=(\omega, {\bf k})$ is the four-momentum of photon, $k^2=0$,
$x(t)=(t, {\bf r}(t))$, $R(t)$ is the matrix element depending on particle
spin. To calculate the combination $R^{\ast}(t_2)R(t_1)$
it is convenient to start from Eq.(7.59) of \cite{BKS}. Summing over
the final electron polarization we find that the combination
$R^{\ast}(t_2)R(t_1)$ in the expression for
the probability of radiation (\ref{2.0}) has the form
\begin{equation}
R^{\ast}(t_2)R(t_1) = \frac{\varepsilon}{2\varepsilon'\gamma^2}
\left\{\frac{\omega^2}{\varepsilon\varepsilon'}+
\left(\frac{\varepsilon}{\varepsilon'}+\frac{\varepsilon'}{\varepsilon}
\right){\bf p}{\bf p}'+i\frac{\omega}{\varepsilon}
(({\bf p}'-{\bf p})\times{\bf v}) \mbox{\boldmath$\zeta$} \right\},
\label{2.15}
\end{equation}
where $\mbox{\boldmath$\zeta$}$ is the vector describing the
initial polarization of the electron (in its rest frame), ${\bf
p}= \gamma \mbox{\boldmath$\vartheta$}(t_1)$ and ${\bf p}'=\gamma
\mbox{\boldmath$\vartheta$}(t_2)$. The first two terms coincide
with Eq.(7.60) of \cite{BKS} (unpolarized electrons), the third
term depends on electron polarization.

In the paper \cite{BK1}(see also \cite{BK2}) we derived the
following expression for the spectral distribution of the
probability of radiation per unit time for unpolarized electron
\begin{eqnarray}
&& \frac{dW}{d\omega}=\frac{2\alpha m^2}{\varepsilon^2} {\rm Im}T,
\nonumber \\
&& T=\left<0|R_1\left(G^{-1}-G_0^{-1}\right)+ R_2 {\bf
p}\left(G^{-1}-G_0^{-1}\right){\bf p}|0\right>, \label{2.1}
\end{eqnarray}
where
\begin{eqnarray}
&& \hspace{-4mm} W = \frac{dw}{dt},\quad
R_1=\frac{\omega^2}{\varepsilon\varepsilon'},\quad
R_2=\frac{\varepsilon}{\varepsilon'}+\frac{\varepsilon'}{\varepsilon},
\quad\varepsilon'=\varepsilon-\omega;
\nonumber \\
&& \hspace{-4mm}G_0={\cal H}_0+1,\quad {\cal H}_0={\bf p}^2,\quad
{\bf
p}=-i\mbox{\boldmath$\nabla$}_{\mbox{\boldmath$\varrho$}},\quad
G={\cal H}+1,\quad {\cal H}={\bf
p}^2-iV(\mbox{\boldmath$\varrho$}), \quad
\nonumber \\
&&
\hspace{-4mm}V(\mbox{\boldmath$\varrho$})=Q\mbox{\boldmath$\varrho$}^2
\left(L_1+\ln \frac{4}{\mbox{\boldmath$\varrho$}^2}-2C \right),
\quad Q=\frac{2\pi Z^2\alpha^2\varepsilon \varepsilon'
n_a}{m^4\omega},\quad L_1=\ln \frac{a_{s2}^2}{\lambda_c^2},
\nonumber \\
&& \hspace{-4mm}\frac{a_{s2}}{\lambda_c}=183Z^{-1/3}{\rm
e}^{-f},\quad
f=f(Z\alpha)=(Z\alpha)^2\sum_{k=1}^{\infty}\frac{1}{k(k^2+(Z\alpha)^2)},
\label{2.2}
\end{eqnarray}
here $C=0.577216 \ldots$ is Euler's constant, $n_a$ is the number
density of atoms in the medium, $\mbox{\boldmath$\varrho$}$ is
the coordinate in the two-dimensional space measured in the
Compton wavelength $\lambda_c$, which is conjugate to the space
of the transverse momentum transfers measured in the electron
mass $m$.

The total probability of radiation $W$ is connected with imaginary
part of the radiative correction to the electron mass according to
\begin{equation}
m\Delta m=\varepsilon \Delta \varepsilon,\quad -2{\rm Im}\Delta
\varepsilon=W,\quad {\rm Im}~\Delta m = - \frac{\varepsilon}{2m}
W.
\label{2.3}
\end{equation}
Since the amplitude $T$ is the analytic function of the potential
$V$ we have that
\begin{equation}
\Delta m = -\alpha m \int_{0}^{\varepsilon}
\frac{d\omega}{\varepsilon}T.
\label{2.4}
\end{equation}
One can derive this formula considering the self-energy diagram
and the corresponding amplitude of forward scattering of electron
(see Eqs.(12.18) and (12.39) in \cite{BKF}).
Here it is helpful also to use the approach formulated in \cite{BK3}.

In the presence of a homogeneous external field the Hamiltonian
(\ref{2.2}) acquires the linear over coordinate term. One can find
the explicit form of this term using the Eq.(7.84) of \cite{BKS}
and carrying out the scale transformation of variables according
to Eq.(2.7) of \cite{BK1}:
\begin{equation}
\Delta{\cal H}=\frac{2}{u}\mbox{\boldmath$\chi$}\mbox{\boldmath$\varrho$},
\label{2.5}
\end{equation}
where
\begin{equation}
\mbox{\boldmath$\chi$}=\frac{\varepsilon}{m^3}{\bf F},\quad
{\bf F}=e({\bf E}_{\perp}+{\bf v}\times {\bf H}),\quad
u=\frac{\omega}{\varepsilon'},
\label{2.5a}
\end{equation}
here ${\bf E}_{\perp}$ is the electric field strength transverse
to the velocity of particle ${\bf v}$, ${\bf H}$ is the magnetic
field strength, $\chi=|\mbox{\boldmath$\chi$}|$ is the known parameter
characterizing quantum effects in the radiation process.

We split the potential $V(\mbox{\boldmath$\varrho$})$ in the same
way as in Eq.(2.17) of \cite{BK1}:
\begin{eqnarray}
&& V(\mbox{\boldmath$\varrho$})=V_c(\mbox{\boldmath$\varrho$})
+v(\mbox{\boldmath$\varrho$}),\quad V_c(\mbox{\boldmath$\varrho$})=
q\mbox{\boldmath$\varrho$}^2, \quad q=QL_c,
\nonumber \\
&&L_c \equiv L(\varrho_c)
=\ln \frac{a_{s2}^2}{\lambda_c^2 \varrho_c^2},\quad
v(\mbox{\boldmath$\varrho$})=-\frac{q\mbox{\boldmath$\varrho$}^2}{L_c}
\left(\ln \frac{\mbox{\boldmath$\varrho$}^2}{4\varrho_c^2}+2C \right).
\label{2.6}
\end{eqnarray}
The definition of the parameter $\varrho_c$ will be considered below.
According to this splitting and taking into account
the addition to the Hamiltonian (\ref{2.5}) we present the propagators in
Eq.(\ref{2.1}) as
\begin{equation}
G^{-1}-G_0^{-1} = G^{-1} - G_F^{-1}+ G_F^{-1} -G_0^{-1},
\label{2.8}
\end{equation}
where
\begin{eqnarray}
&& G_F= {\cal H}_F+1,\quad G=G_F-iv,
\nonumber \\
&& {\cal H}_F = {\bf p}^2+V_F,\quad V_F=-iV_c
+\frac{2}{u}\mbox{\boldmath$\chi$}\mbox{\boldmath$\varrho$}.
\label{2.9}
\end{eqnarray}
The representation of the propagator $G^{-1}$ permits to carry out
its decomposition over the "perturbation" $v$
\begin{equation}
G^{-1} - G_F^{-1} = G_F^{-1} iv G_F^{-1} + G_F^{-1} iv G_F^{-1}
iv G_F^{-1} + \ldots
\label{2.10}
\end{equation}
The procedure of matrix elements calculation in this decomposition
was formulated in \cite{BK1}, see Eqs.(2.30), (2.31). Here the
basic matrix element is \newline
$<\mbox{\boldmath$\varrho$}_1|G_F^{-1}|\mbox{\boldmath$\varrho$}_2>$.
The matrix element for the case when an external field is absent
was calculated in \cite{BK1}, Eqs.(2.20)-(2.27):
\begin{eqnarray}
\hspace{-12mm}&& <\mbox{\boldmath$\varrho$}_1|G_c^{-1}|\mbox{\boldmath$\varrho$}_2>
=i\int_{0}^{\infty}dt \exp(-it)
K_c(\mbox{\boldmath$\varrho$}_1, \mbox{\boldmath$\varrho$}_2, t),\quad
G_c= {\bf p}^2 +1 -iq\mbox{\boldmath$\varrho$}^2,
\nonumber \\
\hspace{-12mm}&&K_c(\mbox{\boldmath$\varrho$}_1, \mbox{\boldmath$\varrho$}_2, t)=
\frac{\nu}{4\pi i \sinh \nu t} \exp \left\{ \frac{i\nu}{4}
\left[ (\mbox{\boldmath$\varrho$}_1^2+\mbox{\boldmath$\varrho$}_2^2)
\coth \nu t - \frac{2}{\sinh \nu t}
\mbox{\boldmath$\varrho$}_1\mbox{\boldmath$\varrho$}_2\right] \right\},
\label{2.11}
\end{eqnarray}
where $\nu=2\sqrt{iq}$.

The matrix element of propagator $G_F^{-1}$ can be obtained from
(\ref{2.11}) by transformation of the Hamiltonian ${\cal H}_F$ to
the quadratic form over coordinate
\begin{equation}
{\cal H}_F={\bf p}^2 -iq\mbox{\boldmath$\varrho$}^2+
\frac{2}{u}\mbox{\boldmath$\chi$}\mbox{\boldmath$\varrho$}= {\bf p
}^2
-iq\mbox{\boldmath$\varrho$}'^{2}-i\frac{\mbox{\boldmath$\chi$}^2}{qu^2},\quad
\mbox{\boldmath$\varrho$}'=\mbox{\boldmath$\varrho$}+i\frac{\mbox{\boldmath$\chi$}}{qu}
\label{2.12}
\end{equation}
In the new variables the Hamiltonian (\ref{2.12}) has the form
similar to (\ref{2.9}) and one can use (\ref{2.11}). Substituting
into (\ref{2.11}) $\mbox{\boldmath$\varrho$}_{1,2}\rightarrow
\mbox{\boldmath$\varrho$}'_{1,2}$ and taking into account the
constant term in the Hamiltonian (\ref{2.12}) we obtain
\begin{eqnarray}
\hspace{-16mm}&& K_F(\mbox{\boldmath$\varrho$}_1,
\mbox{\boldmath$\varrho$}_2, t)=K_c(\mbox{\boldmath$\varrho$}_1,
\mbox{\boldmath$\varrho$}_2,
t)K_{\chi}(\mbox{\boldmath$\varrho$}_1,
\mbox{\boldmath$\varrho$}_2, t)
\nonumber \\
\hspace{-16mm}&&K_{\chi}(\mbox{\boldmath$\varrho$}_1,
\mbox{\boldmath$\varrho$}_2,
t)=\exp\left[-\frac{4i\mbox{\boldmath$\chi$}^2t}{u^2\nu^2}\left(
1-\frac{2}{\nu t}\tanh \frac{\nu
t}{2}\right)-\frac{2i}{u\nu}\mbox{\boldmath$\chi$}(\mbox{\boldmath$\varrho$}_1+
\mbox{\boldmath$\varrho$}_2)\tanh \frac{\nu t}{2}\right].
\label{2.13}
\end{eqnarray}

The presence in the problem under consideration the axial vector
\begin{equation}
{\bf H}_R=\gamma({\bf H}_{\perp}+{\bf E}\times {\bf v})
\label{2.14}
\end{equation}
leads to appearance in Eqs.(\ref{2.1}) and (\ref{2.4}) additional
terms depending on the electron polarization
$\mbox{\boldmath$\zeta$}$ (see Eq.(\ref{2.15})). According to
analysis performed in Subsec.7.4 of \cite{BKS} and in \cite{BK1}
the vector ${\bf p}'$ in (\ref{2.15}) gets over to the operator
${\bf p}$ standing in the formula (\ref{2.1}) for $T$ from the
right of the propagator $G^{-1}$ and the vector ${\bf p}$ in
(\ref{2.15}) gets over to the operator ${\bf p}$ standing in the
formula (\ref{2.1}) for $T$ from the left of the propagator
$G^{-1}$. As a result we have for the addition to $T$ in
Eqs.(\ref{2.1}) and (\ref{2.4}) depending on the spin vector
$\mbox{\boldmath$\zeta$}$:
\begin{equation}
T \rightarrow T + T_{\mbox{\boldmath$\zeta$}},\quad
T_{\mbox{\boldmath$\zeta$}}=i\frac{\omega}{\varepsilon} \left(
<0|(G^{-1}{\bf p}-{\bf p}G^{-1})|0>\times{\bf v} \right)
\mbox{\boldmath$\zeta$}.
\label{2.16}
\end{equation}
For the radiative correction to the electron mass
we have respectively
\begin{equation}
\Delta m \rightarrow \Delta M =
\Delta m + \Delta m_{\mbox{\boldmath$\zeta$}}.
\label{2.16a}
\end{equation}
It should be noted that expressions for $T$ (\ref{2.1}) and
(\ref{2.16}) have universal form while the specific of the
particle motion is contained in the propagator $G^{-1}$ through
the effective potential
\begin{equation}
V_{eff}(\mbox{\boldmath$\varrho$})=-iV(\mbox{\boldmath$\varrho$})
+\frac{2}{u}\mbox{\boldmath$\chi$}\mbox{\boldmath$\varrho$}.
\label{2.17}
\end{equation}

In the present paper we restrict ourselves to the main term
in the decomposition (\ref{2.8}). This means that result will have
the logarithmic accuracy over the scattering (but not over an external
field). With regard for an external field the parameter $\varrho_c$
in Eq.(\ref{2.6}) is defined by a set of equations:
\begin{eqnarray}
&& \varrho_c =1\quad {\rm for}\quad 4\frac{\chi^2}{u^2}+4QL_1 \leq 1;
\nonumber \\
&& 4Q\varrho_c^4 \left[4\frac{\chi^2}{u^2}+|\nu(\varrho_c)|^2 \right]=1
\quad {\rm for} \quad 4\frac{\chi^2}{u^2}+ 4QL_1 \geq 1,
\label{2.18}
\end{eqnarray}
where $\nu=2\sqrt{iq}$, $q=QL_c$, $L_1$ and $Q$ are defined in
Eq.(\ref{2.2}), $L_c$ is defined in Eq.(\ref{2.6}).

The matrix elements entering into the mass correction
have in the used approximation the following form
\begin{eqnarray}
&& \left<0|G_F^{-1}-G_0^{-1} |0\right>
=\frac{1}{4\pi}\int_{0}^{\infty}\exp(-it)\left(\frac{\nu}{\sinh \nu t}\varphi
-\frac{1}{t} \right)dt,
\nonumber \\
&& \left<0|{\bf p}(G_F^{-1}-G_0^{-1}){\bf p} |0\right>
=-\frac{1}{4\pi}\int_{0}^{\infty}\exp(-it)\Bigg[\frac{\nu}{\sinh \nu t}
\Bigg(\frac{4\chi^2}{u^2\nu^2}\tanh^2\frac{\nu t}{2}
\nonumber \\
&& +\frac{i\nu}{\sinh \nu t} \Bigg)\varphi -\frac{i}{t^2}
\Bigg]dt,\quad \left<0|(G_F^{-1}{\bf p}-{\bf p}G_F^{-1}) |0\right>
=\frac{\mbox{\boldmath$\chi$}}{2\pi u}
\int_{0}^{\infty}\exp(-it)\frac{\varphi}{\cosh^2 \frac{\nu
t}{2}}dt,
\nonumber \\
&& \varphi \equiv \varphi(\chi, \nu, t)
=\exp\left[-\frac{4i\chi^2t}{\nu^2 u^2}\left( 1-\frac{2}{\nu
t}\tanh\frac{\nu t}{2} \right) \right].
\label{2.19}
\end{eqnarray}
The two first equations are consistent with obtained in
\cite{BKS2}. Thus, in the used approximation ($G=G_F$) we have for
$T_{\mbox{\boldmath$\zeta$}}$ Eq.(\ref{2.16})
\begin{equation}
T_{\mbox{\boldmath$\zeta$}}=\frac{i}{2\pi u}\frac{\omega}{\varepsilon}
\int_{0}^{\infty}\exp(-it)\frac{\varphi}{\cosh^2 \frac{\nu t}{2}}dt
(\mbox{\boldmath$\zeta$}\mbox{\boldmath$\chi$}{\bf v}),
\label{2.20}
\end{equation}
where $(\mbox{\boldmath$\zeta$}\mbox{\boldmath$\chi$}{\bf v})=
(\mbox{\boldmath$\zeta$}\cdot(\mbox{\boldmath$\chi$}\times{\bf v}))$.

\section{Anomalous magnetic moment of electron}
\setcounter{equation}{0}

Within relativistic accuracy (up to terms of higher order over
$1/\gamma$) the combination of vectors entering in
Eq.(\ref{2.20}) can be written as (see Eq.(12.19) in \cite{BKF}
and \cite{BKS3})
\begin{equation}
(\mbox{\boldmath$\zeta$}\mbox{\boldmath$\chi$}{\bf v})=
2\mu_0\frac{\mbox{\boldmath$\zeta$}{\bf H}_R}{m},
\label{3.1}
\end{equation}
where ${\bf H}_R$ defined in Eq.(\ref{2.14}) is the magnetic field
in the electron rest frame. Here we use that
$\mbox{\boldmath$\chi$}=\chi{\bf s}$, where ${\bf s}$ is the unit
vector in the direction of acceleration (this vector is used in
\cite{BKF}).

In the electron rest frame one can consider the value Re~$\Delta
m_{\mbox{\boldmath$\zeta$}}$ depending on the electron spin as the
energy of interaction of the AMM of electron with the magnetic
field ${\bf H}_R$ (see Eq.(12.23) in \cite{BKF} and  \cite{BKS3})
\begin{equation}
{\rm Re}~\Delta m_{\mbox{\boldmath$\zeta$}}=-\mu'
\mbox{\boldmath$\zeta$}{\bf H}_R,
\label{3.2}
\end{equation}

Taking into account Eqs.(\ref{2.4}), (\ref{2.16}), (\ref{2.16a}),
(\ref{2.20}), (\ref{3.1}), (\ref{3.2}) we obtain the following
general expression for the AMM of electron moving in a medium in
the presence of an external electromagnetic field:
\begin{equation}
\frac{\mu'}{\mu_0}=-\frac{\alpha}{\pi} {\rm Im}~
\int_{0}^{\infty} \frac{du}{(1+u)^3}
\int_{0}^{\infty}\exp(-it)\frac{\varphi}{\cosh^2 \frac{\nu t}{2}}dt
\label{3.3}
\end{equation}

In the absence of scattering ($\nu \rightarrow 0$) the expression
(\ref{3.3}) gets over into the formula for the AMM of electron in external
field (see Eq.(12.24) in \cite{BKF} and \cite{R}, \cite{BKS3})
\begin{equation}
\frac{\mu'}{\mu_0}=-\frac{\alpha}{\pi} {\rm Im}~
\int_{0}^{\infty} \frac{du}{(1+u)^3}
\int_{0}^{\infty}
\exp\left[-it\left(1+\frac{\chi^2 t^2}{3u^2} \right)\right]dt
\label{3.4}
\end{equation}

In the weak external field ($\chi \ll 1, \varphi \simeq 1$) we
obtain the formula for the AMM of electron under influence of
multiple scattering
\begin{equation}
\frac{\mu'}{\mu_0}=-\frac{\alpha}{\pi} {\rm Im}~
\int_{0}^{\infty} \frac{du}{(1+u)^3}
\int_{0}^{\infty}\exp(-it)\frac{1}{\cosh^2 \frac{\nu
t}{2}}dt=\frac{\alpha}{2\pi}r,
\label{3.5}
\end{equation}
where
\begin{eqnarray}
&& r={\rm Re}J,\quad J=2i\int_{0}^{\infty} \frac{du}{(1+u)^3}
\int_{0}^{\infty}\exp(-it)\frac{1}{\cosh^2 \frac{\nu t}{2}}dt
\nonumber \\
&& =4i\int_{0}^{\infty} \frac{du}{(1+u)^3}
\frac{1}{\nu}\left[\frac{2i}{\nu}\beta\left(\frac{i}{\nu} \right)-1 \right],
\nonumber \\
&& \beta=\frac{1}{2}\left[\psi\left(\frac{1+x}{2} \right)-
\psi\left(\frac{x}{2} \right) \right],
\label{3.6}
\end{eqnarray}
where $\psi(x)$ is the logarithmic derivative of the gamma function.

The dependence of the AMM of electron on its energy $\varepsilon$ in gold
is shown in Fig. It is seen that at energy $\varepsilon=500$~GeV
value of AMM is 0.85 part of standard quantity of AMM (SQ), at energy
$\varepsilon=1$~TeV it is 0.77 part of SQ and at energy
$\varepsilon=5.5$~TeV it is 0.5 part of SQ.
Actually in all heavy elements the behavior of AMM of
electron will be quite similar, i.e. the scale of energy where the
AMM of electron deviates from SQ is of order of TeV.

In the case of weak effect of multiple scattering
\begin{equation}
\nu^2=\frac{i}{u}\nu_b^2,\quad
\nu_b^2=\frac{\varepsilon}{\varepsilon_e},\quad
\varepsilon_e=m\left(8\pi
Z^2\alpha^2n_a\lambda_c^3L_1\right)^{-1},
\label{3.7}
\end{equation}
where $\varepsilon_e$ is the characteristic electron energy
starting with the multiple scattering distorted the whole
spectrum of radiation ($\varepsilon_e=2.6$~TeV in gold and it has a
similar value for heavy elements).
Integrating in Eq.(\ref{3.6}) by parts over
the variable $t$ we find
\begin{equation}
J=1+f,\quad f=f(\nu_b)= -2\int_{0}^{\infty} \frac{du}{(1+u)^3}
\int_{0}^{\infty}\exp(-it)\frac{\tanh \frac{\nu t}{2}}{\cosh^2
\frac{\nu t}{2}}dt. \label{3.8}
\end{equation}
The asymptotic behavior of the function $f$ at $\nu_b \ll
1~(\varepsilon \ll \varepsilon_e)$ calculated in Appendix:
\begin{equation}
f(\nu_b)= \nu_b^2\left[-\frac{\pi}{2}+2i\left(\ln \frac{2}{\nu_b}
+C-\frac{7}{4}+A \right)\right],
\label{3.9}
\end{equation}
where
\begin{equation}
A=\int_{0}^{\infty}\left(3 \ln t \tanh t +\frac{\sinh t
-t}{t^2}\right)\frac{dt}{\cosh^3t} = -0.364291\ldots
\label{3.10}
\end{equation}
Substituting this result into (\ref{3.5}) we obtain for AMM of
electron at $\varepsilon \ll \varepsilon_e$
\begin{equation}
\frac{\mu'}{\mu_0}=\frac{\alpha}{2\pi}\left(1-
\frac{\pi}{2}\frac{\varepsilon}{\varepsilon_e}\right).
\label{3.11}
\end{equation}

At very high energy $\varepsilon \gg \varepsilon_e$
the effect of multiple scattering becomes strong. In this case
(see Eq.(3.3) in \cite{BK2})
\begin{equation}
\nu^2=\frac{i\varepsilon}{u \varepsilon_e}\left(1+\frac{1}{2L_1}
\ln\frac{\varepsilon}{\varepsilon_e u} \right),
\label{3.12}
\end{equation}
the main contribution into the integral (\ref{3.6}) give the
regions $u \sim 1$ and $t \sim 1/|\nu| \ll 1$. Making the
substitution of variable as in (\ref{a.8}) and expanding the
exponent as in Appendix we have from (\ref{a.11}) the asymptotic
expansion of AMM of electron:
\begin{equation}
r=\frac{2\pi}{\alpha} \frac{\mu'}{\mu_0}=\frac{\pi}{2\sqrt{2}}
\sqrt{\frac{\varepsilon_c}{\varepsilon}}\left(1-
\frac{\pi^2}{2}\frac{\varepsilon_c}{\varepsilon}\right), \quad
\varepsilon_c =\varepsilon_e\frac{L_1}{L_0},\quad
L_0=L_1+\frac{1}{2}\ln \frac{\varepsilon}{\varepsilon_e}.
\label{3.13}
\end{equation}
Redefinition of the characteristic energy $\varepsilon_e \rightarrow
\varepsilon_c$ is connected with enlargement of the radiation cone
(in comparison with $1/\gamma$) or in another terms with increasing
of the characteristic transverse momentum transfers due to the multiple
scattering.

\section{Spin correction to the intensity of radiation}
\setcounter{equation}{0}

To obtain the additional term in the radiation intensity depending
on the spin of the radiating electron one has to substitute the formula
(\ref{2.20}) into expression for the spectral distribution of the
probability of radiation per unit time Eq.(\ref{2.1}) and to multiply it
on the photon energy $\omega$. As a result we obtain the mentioned term
\begin{equation}
I_{\mbox{\boldmath$\zeta$}}=
(\mbox{\boldmath$\zeta$}\mbox{\boldmath$\chi$}{\bf v})
\frac{\alpha m^2}{\pi} {\rm Re} \int_{0}^{\infty}
\frac{udu}{(1+u)^4}
\int_{0}^{\infty}\exp(-it)\frac{\varphi}{\cosh^2 \frac{\nu t}{2}}dt.
\label{4.1}
\end{equation}
So for polarized electrons the total intensity of radiation
is the sum of Eq.(3.8) of \cite{BK2} and
$I_{\mbox{\boldmath$\zeta$}} L_{rad}^0/\varepsilon$ ({\ref{4.1}}).

At $\varepsilon \ll \varepsilon_e$ one can decompose
$\cosh^{-2} \nu t/2 \simeq 1- \nu^2 t^2/4$ in integral over $t$ in
(\ref{4.1}). We find
\begin{equation}
I_{\mbox{\boldmath$\zeta$}} \simeq
(\mbox{\boldmath$\zeta$}\mbox{\boldmath$\chi$}{\bf v})
\frac{\alpha m^2}{6\pi} \frac{\varepsilon}{\varepsilon_e},\quad
\frac{I_{\mbox{\boldmath$\zeta$}}}{I} \simeq \frac{2}{3}
(\mbox{\boldmath$\zeta$}\mbox{\boldmath$\chi$}{\bf v}),
\label{4.2}
\end{equation}
Where $I$ is given by Eq.(3.9) of \cite{BK2}.

In the opposite case of large energies $\varepsilon \gg \varepsilon_e$
the main contribution into integrals in ({\ref{4.1}}) gives the region
$u \sim 1,~t \ll 1$. Then substituting the exponent by unity we find
\begin{eqnarray}
&& I_{\mbox{\boldmath$\zeta$}} \simeq
(\mbox{\boldmath$\zeta$}\mbox{\boldmath$\chi$}{\bf v})
\frac{\alpha m^2}{\pi}\sqrt{\frac{2\varepsilon_c}{\varepsilon}}
\int_{0}^{\infty} \int_{0}^{\infty}\frac{u^{3/2}du}{(1+u)^4}
=(\mbox{\boldmath$\zeta$}\mbox{\boldmath$\chi$}{\bf v})
\frac{\alpha m^2}{8\sqrt{2}}\sqrt{\frac{\varepsilon_c}{\varepsilon}}
\nonumber \\
&& \frac{I_{\mbox{\boldmath$\zeta$}}}{I} \simeq \frac{4}{9}
\frac{\varepsilon_c}{\varepsilon}
(\mbox{\boldmath$\zeta$}\mbox{\boldmath$\chi$}{\bf v}),
\label{4.3}
\end{eqnarray}
where $I$ is given by Eq.(3.11) of \cite{BK2}.

\section{Conclusion}
\setcounter{equation}{0}

Here we discuss the possibility of experimental observation of
influence of medium on the AMM of electron. At a very high energy
where the observation becomes feasible the rotation angle $\varphi$
of the spin vector $\mbox{\boldmath$\zeta$}$ in the transverse
to the particle velocity ${\bf v}$ magnetic field ${\bf H}$ depends
only on the value of AMM and doesn't depend on the particle
energy:
\begin{equation}
\varphi =\left(\frac{m}{\varepsilon}+ \frac{\mu'}{\mu_0}
\right)\frac{e H}{m}l \simeq
r\frac{\alpha}{2\pi}\frac{H}{H_0}\frac{l}{\lambda_c}, \label{5.1}
\end{equation}
where $l$ is the path of electron in the field, $H_0=m^2/e=4.41
\cdot 10^{13}$~Oe. The dependence of $r$ on energy $\varepsilon$ is
found in Sec.3 and shown in Fig.

Since at radiation of hard photons in a medium the picture is
quite complicated:energy losses, cascade processes, spin flip and
depolarization, it is desirable to measure the particles which
don't radiate photons on the path $l$. The number of such
particles $N$ is determined by the total probability of radiation
in a medium found in \cite{BK2}, Eqs.(3.12)-(3.14):
\begin{eqnarray}
&& N=N_0\exp(-\psi(\varepsilon)),\quad
\psi(\varepsilon)=W(\varepsilon)l=\frac{k(\varepsilon)\varphi_{SQ}}
{2\chi(\varepsilon_e)};\quad
k(\varepsilon)=W(\varepsilon)L_{rad}^0,
\nonumber \\
&& (L_{rad}^0)^{-1}
=\frac{\alpha}{4\pi}\frac{m^2}{\varepsilon_e},\quad
\chi(\varepsilon_e)=\frac{\varepsilon_e}{m}\frac{H}{H_0},
\label{5.2}
\end{eqnarray}
where $\varphi_{SQ}$ is the rotation angle for standard value of
AMM in QED ($r=1$), $N_0$ is the number of initial electrons. With
energy increase the function $k(\varepsilon)$ decreases \cite{BK2}
\begin{equation}
k(\varepsilon \ll \varepsilon_e) \simeq \frac{4}{3}\left( \ln
\frac{\varepsilon_e}{\varepsilon}+1.96 \right),\quad
k(\varepsilon = \varepsilon_e) \simeq 3.56, \quad k(\varepsilon
\gg \varepsilon_e) \simeq
\frac{11\pi}{4\sqrt{2}}\sqrt{\frac{\varepsilon_e}{\varepsilon}}.
 \label{5.3}
\end{equation}
The crucial part of the experiment is an accuracy of measurement
of electron polarization before target and after target. If one
supposes that spin rotation angle can be measured with accuracy
noticeably better than 1/10 then we can put that
$(1-r)\varphi_{SQ}=1/10$. In the gold we find for the energy
$\varepsilon=\varepsilon_e=2.61~$TeV and the magnetic field $H=4
\cdot 10^{5}$Oe that $1-r = 0.371$, the path of electron in the
target is $l \simeq 1~$cm and number of electron traversing the
target without energy loss is $N \simeq 3.2 \cdot 10^{-5} N_0$.
This estimates show that the measurement of the effect found in
this paper will be feasible in the not very distant future.

{\bf Acknowledgments}
\vspace{0.2cm}
This work
was supported in part by the Russian Fund of Basic Research under Grant
00-02-18007.

\setcounter{equation}{0}
\Alph{equation}

\appendix

\section{Appendix}

In the case $\nu_b \ll 1$ (\ref{3.7}) we present the function
$f(\nu)$ (\ref{3.8}) as
\begin{eqnarray}
&& f(\tilde{\nu})=f_1(\tilde{\nu})+f_2(\tilde{\nu}),\quad
\nu=\frac{\tilde{\nu}}{\sqrt{u}},
\nonumber \\
&& f_1=-2\int_{0}^{\infty}\nu
\frac{du}{(1+u)^3}\int_{0}^{\infty}\displaystyle{\frac{\exp(-it)}{\cosh^3\frac{\nu
t}{2}}}\left(\sinh\frac{\nu t}{2}-\frac{\nu t}{2}\right)dt,
\nonumber \\
&& f_2 =-\int_{0}^{\infty}\nu^2
\frac{du}{(1+u)^3}\int_{0}^{\infty}\frac{\exp(-it)}{\cosh^3\frac{\nu
t}{2}}dt.
\label{a.1}
\end{eqnarray}
In the integral in $f_1$ the main contribution give $u \sim
\nu_b^2 \ll 1$ so that
\begin{equation}
f_1=-2\tilde{\nu}\int_{0}^{\infty}
\frac{du}{\sqrt{u}}\int_{0}^{\infty}\displaystyle{\frac{\exp(-it)}{\cosh^3\frac{\tilde{\nu}
t}{2\sqrt{u}}}}\left(\sinh\frac{\tilde{\nu}
t}{2\sqrt{u}}-\frac{\tilde{\nu} t}{2\sqrt{u}}\right)dt.
\label{a.2}
\end{equation}
Substituting variables $x=\sqrt{u}$ and $z=\nu t/2x$ we have
\begin{equation}
f_1=2\tilde{\nu}^2\int_{0}^{\infty} \frac{1}{z^2} \frac{\sinh
z-z}{\cosh^3z}dz. \label{a.3}
\end{equation}
The function $f_2$ we present as
\begin{eqnarray}
\hspace{-12mm}&&
f=f_2^{(1)}+f_2^{(2)},~f_2^{(1)}=\tilde{\nu}^2\int_{0}^{\infty}
\frac{du}{u}\left(\frac{1}{1+u}-\frac{1}{(1+u)^3}\right)
\int_{0}^{\infty}\displaystyle{\frac{\exp(-it)}{\cosh^3\frac{\tilde{\nu}
t}{2\sqrt{u}}}}t dt
\nonumber \\
\hspace{-12mm}&& \simeq
-\tilde{\nu}^2\int_{0}^{\infty}\frac{2+u}{(1+u)^3}du =
-\frac{3}{2}\tilde{\nu}^2,
\nonumber \\
\hspace{-12mm}&& f_2^{(2)} =-\tilde{\nu}^2\int_{0}^{\infty}
\frac{du}{u(1+u)}\int_{0}^{\infty}\frac{\exp(-it)}{\cosh^3\frac{\tilde{\nu}
t}{2\sqrt{u}}}t dt=-\tilde{\nu}^2\int_{0}^{\infty}\frac{dx}{1+x}
\int_{0}^{\infty}\displaystyle{\frac{\exp(-it)}{\cosh^3\frac{\tilde{\nu}
t}{2}\sqrt{x}}}t dt.
\label{a.4}
\end{eqnarray}
Introducing in the last integral the variable
$z=\tilde{\nu}^2t^2x/4$ we find
\begin{equation}
f_2^{(2)} =-\tilde{\nu}^2\int_{0}^{\infty}\exp(-it)t dt
\int_{0}^{\infty}
\frac{dz}{\left(z+\tilde{\nu}^2t^2/4\right)\cosh^3\sqrt{z}}
\label{a.5}
\end{equation}
Integrating in (\ref{a.5}) by parts over the variable $z$ we find
\begin{eqnarray}
\hspace{-12mm}&&
f_2^{(2)} \simeq -\tilde{\nu}^2
\int_{0}^{\infty}\exp(-it) \left(-\ln \frac{\tilde{\nu}^2 t^2}{4}
+ \int_{0}^{\infty} \frac{3}{2\sqrt{z}} \ln z \frac{\sinh \sqrt{z}}
{\cosh^4\sqrt{z} }dz \right) t dt
\nonumber \\
\hspace{-12mm}&&
=2\tilde{\nu}^2\left(\frac{i\pi}{2}-\psi(2)-\ln \frac{\tilde{\nu}}{2}+
3\int_{0}^{\infty}\ln x \frac{\tanh x}{\cosh^3x}dx \right).
\label{a.6}
\end{eqnarray}
Combining the results (\ref{a.3}), (\ref{a.4}) and (\ref{a.7})
we obtain
\begin{eqnarray}
\hspace{-12mm}&&
f(\tilde{\nu}) = 2\tilde{\nu}^2
\left(\frac{i\pi}{2}+\ln \frac{2}{\tilde{\nu}}-\frac{7}{4}+C+A \right),
\nonumber \\
\hspace{-12mm}&&
A=\int_{0}^{\infty} \frac{1}{\cosh^3x}\left(\frac{\sinh x -x}{x^2}
+3 \ln x \tanh x \right) dx.
\label{a.7}
\end{eqnarray}

In the case $|\tilde{\nu}| \gg 1$ we present the integral in
Eq.(\ref{3.6}) as
\begin{equation}
J(\tilde{\nu})  =\frac{4i}{\tilde{\nu}}\int_{0}^{\infty}
\frac{\sqrt{u}du}{(1+u)^3}\int_{0}^{\infty}
\frac{dt}{\cosh^2t}\exp\left(-i\frac{2\sqrt{u}t}{\tilde{\nu}}\right).
\label{a.8}
\end{equation}
To calculate the first three terms of the decomposition of
$J(\tilde{\nu})$ over $1/\tilde{\nu}$ one can expands the
exponent in the last integral
\begin{equation}
\exp\left(-i\frac{2\sqrt{u}t}{\tilde{\nu}}\right)=1-\frac{2i\sqrt{u}t}{\tilde{\nu}}-
\frac{2ut^2}{\tilde{\nu}^2}.
\label{a.9}
\end{equation}
The next term of this decomposition contains the logarithmic
divergence of the integral over $u$ and to calculate the term of
the order $1/\tilde{\nu}^3$ one has to use another approach. Using
the known integrals
\begin{eqnarray}
\hspace{-12mm}&&  \int_{0}^{\infty} \frac{t dt}{\cosh^2t}=\ln
2,\quad \int_{0}^{\infty}\frac{t^2dt}{\cosh^2t}=\frac{\pi^2}{12},
\nonumber \\
\hspace{-12mm}&& \int_{0}^{\infty} \frac{\sqrt{u}du}{(1+u)^3}
=\frac{\pi}{8},\quad \int_{0}^{\infty}
\frac{u^{3/2}du}{(1+u)^3}=\frac{3\pi}{8},
\label{a.10}
\end{eqnarray}
we obtain at $|\tilde{\nu}|\gg
1(\tilde{\nu}=\sqrt{i}|\tilde{\nu}|$)
\begin{eqnarray}
\hspace{-12mm}&&  J(\tilde{\nu}) \simeq \frac{4i}{\tilde{\nu}}
\left[\frac{\pi}{8}-\frac{i}{\tilde{\nu}}\ln
2-\frac{\pi^3}{16\tilde{\nu}^2}\right],\quad {\rm
Re}J(\tilde{\nu}) \simeq
\frac{\pi}{2\sqrt{2}|\tilde{\nu}|}\left(1-\frac{\pi^2}{2|\tilde{\nu}|^2}\right),
\nonumber \\
\hspace{-12mm}&& {\rm Im}J(\tilde{\nu}) \simeq
\frac{\pi}{2\sqrt{2}|\tilde{\nu}|} \left(1-\frac{8\sqrt{2}\ln 2
}{\pi |\tilde{\nu}|}+\frac{\pi^2}{2|\tilde{\nu}|^2}\right).
\label{a.11}
\end{eqnarray}

\newpage

{\bf Figure captions}

\vspace{15mm} Anomalous magnetic moment(AMM) of electron in units
$\alpha/2\pi$ ($r$ in Eq.(\ref{3.6}))~in gold vs electron energy
in TeV.

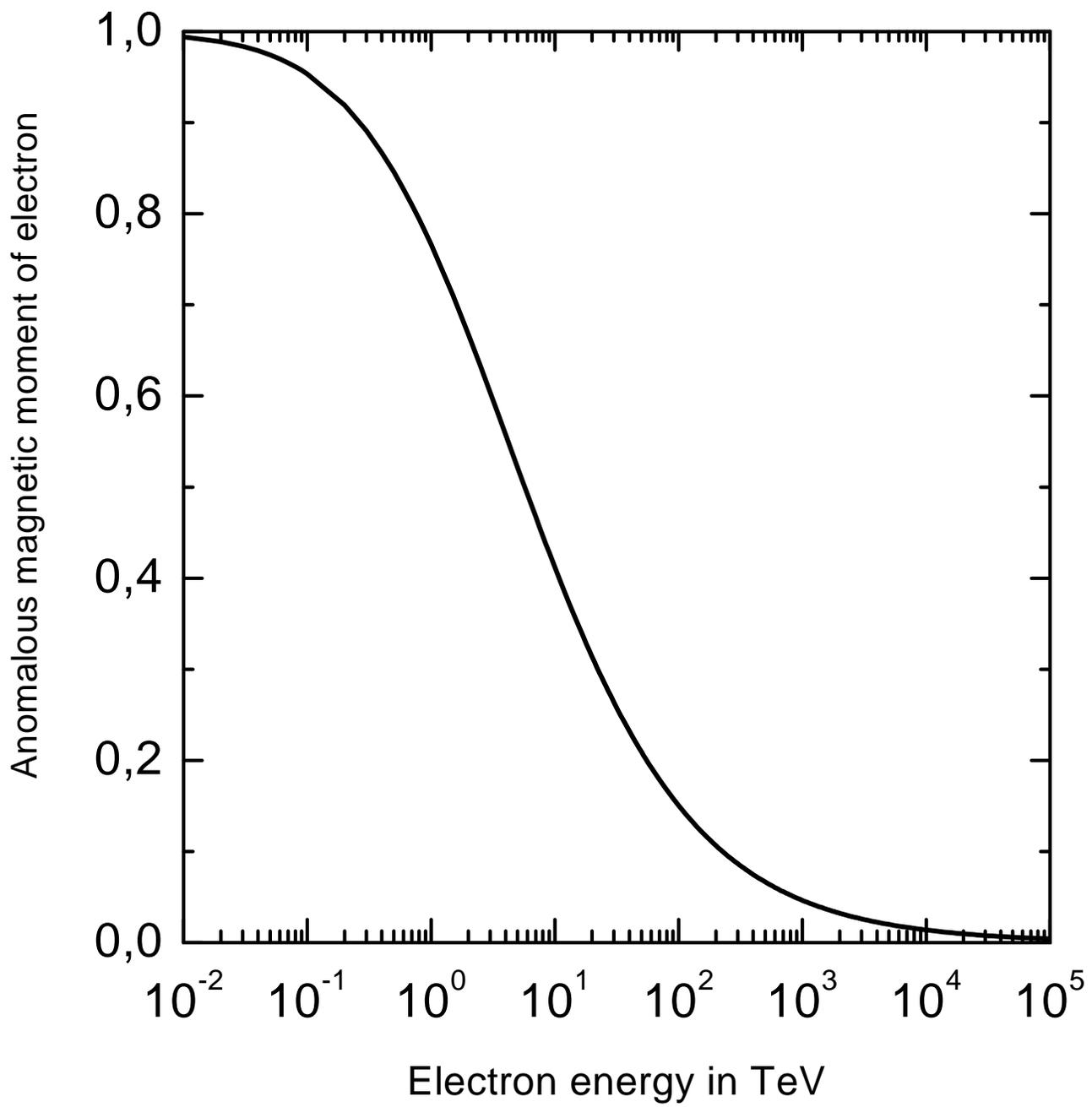
\end{document}